\documentclass[12pt]{iopart}
\usepackage{indentfirst} %%%% This package causes indentation of the first line after \section command
\usepackage{bm} %%%%% This package id used for typesetting bold italic math. letters and symbols
\usepackage{epsfig}
\usepackage{graphicx}
\def\be{\begin{equation}}
\def\ee{\end{equation}}
\def\bea{\begin{eqnarray}}
\def\eea{\end{eqnarray}}

\def\bearst{\begin{eqnarray*}}
\def\eearst{\end{eqnarray*}}

\begin{document}

\title[Solid on Solid Model for Surface Growth in $2+1$ Dimensions]{Solid on Solid Model for Surface Growth in $2+1$ Dimensions}

\author{S. Hosseinabadi$^1$, A. A. Masoudi$^{1,2} $ and  M. Sadegh Movahed$^{3}$ }
\address{$^{1}$Department of Physics, Alzahra University, P.O.Box
19938, Tehran 91167, Iran}
\address{$^{2}$Department  of Applied Mathematics , University of Waterloo , Waterloo, ON N2L 3G1, Canada}
\address{$^{3}$Department of Physics, Shahid Beheshti University, G.C., Evin, Tehran 19839, Iran}

%\ead{m.s.movahed@ipm.ir}
\ead{amasoudi@math.uwaterloo.ca}
\begin{abstract}
We analyze in detail the Solid-On-Solid model (SOS) for growth
processes on a square substrate in $2+1$ dimensions. By using the
Markovian surface properties, we introduce an alternative approach for
determining the roughness exponent of a special type of SOS model-the Restricted-Solid-On-Solid model (RSOS)- in $2+1$ dimensions.
This model is the SOS model with the additional restriction that the height difference must be $S=1$.
Our numerical results show that
the behaviour of the SOS model in $2+1$ dimensions for approximately $S\geq
S_{\times}\sim 8$ belongs to the two different universality classes:
during the initial time stage, $t< t_{\times}$ it belongs to the
Random-Deposition (RD) class, while for $t_{\times}<t\ll
t_{sat}$ it belongs to the
Kardar-Parisi-Zhang (KPZ) universality class. The crossover time ($t_{\times}$) is related to $S$ via a power law with exponent, $\eta=1.99\pm0.02$ at $1\sigma$
confidence level which is the same as that for $1+1$ dimensions
reported in Ref. \cite{e1}. Using the structure function, we compute
the roughness exponent. In contrast to the growth exponent, the
roughness exponent does not show crossover for different values of $S$. The
scaling exponents of the structure function for fixed values of separation
distance versus $S$ in one and two space dimensions are
$\xi=0.92\pm0.05$ and $\xi=0.86\pm0.05$ at $1\sigma$ confidence
level, respectively.

\end{abstract}

\maketitle

\section{Introduction}

Surface growth processes, especially the formation of thin film deposits, have been studied using
various approaches in complex systems and statistical analysis
\cite{e3,a3,a4,a6,2,3,Barabasi}. The factors which control
surface growth phenomena have immense phase space.
Consequently, to be able to analyze these phenomena one needs to make many assumptions, which can lead to results that are unreliable. Combining insights from computational simulation and simplified analysis will likely give better results. It is well-known that the understanding of phenomena such as
advances of bacterial colonies, electrochemical deposition, flameless
fire fronts and molecular-beam-epitaxial growth is of considerable importance in the control of
many interesting growth processes in industries
\cite{Barabasi,e1,e2}.
%Development of thin film roughness of optoelectronic devices has
%great influence on their optical properties.
The simplest surface growth model is the so called statistical deposition model \cite{Barabasi,4}. Some models
proposed to explore growth surfaces, such as the Family model
\cite{5}, Ballistic Deposition(BD) model \cite{6,7}and the Eden model
\cite{8}, are able to account for many of the properties of some real systems. For example,
the BD and Eden models can accurately simulate vapor
deposition and biological growth. However, these models tend to ignore the microscopic details of the interfaces , and cannot provide accurate scaling exponents.
In addition many fractal features of real systems remain unexplained
 \cite{e2,famili90,11}. To solve these problems, one
should modify the above models.

The Solid-On-Solid model (SOS) is more suitable to describe a real
surface's properties than those models described above\cite{e1,kim89,kim911,sarma}. This growth model does
not exhibit strong corrections to scaling and consequently allows us
to determine accurate values of scaling exponents
\cite{Barabasi,kim89,kim911}. The Restricted-Solid-On-Solid (RSOS) model( a modified version of the SOS model), proposed by Kim et al.
\cite{kim89}, is most important due to its wide applicability,
such as for surface roughening modeling via exothermic catalytic
reactions on the substrate \cite{e1}. Various aspects of the Solid-On-Solid
model for surface growth have been studied: the effect of long-range elastic interactions
\cite{muller05}, growth processes with correlated noise
\cite{margo90}, phase transitions as a function of temperature-like parameters \cite{amar90}, the $(001)$-surface morphology of GaAs annealed at
fixed temperature and pressure, the well explained by annealed
version of the RSOS model \cite{e4,zhd03}. Crossover from random to
correlated regime \cite{a1,a2}, relaxation to steady states
\cite{a7}, distribution of local configurations for finite values of
$S$ \cite{e1}, Markov analysis \cite{24}, the effect of hopping in
various local growth rules on the linear and nonlinear fourth-order
dynamical growth equation \cite{sarma92}, growth model in higher
dimensions \cite{ala92} and, more recently, the growth on  fractal
substrates based on the SOS model  \cite{lee}, has also been addressed in the literature.

As mentioned in many previous studies, it is believed that the RSOS
model belongs to the Kardar-Parisi-Zhang (KPZ) universality class in
the continuum limit \cite{park95,huang98}. The KPZ equation is one of
the most important phenomenological theories in which time evolution
of the interface has been characterized by the height function
$h(\vec{r},t)$ at position $\vec{r}$ and time $t$. The governing
equation is given by \cite{10}: \bea
\frac{\partial{h}(\vec{r},t)}{\partial{t}}=\nu\nabla^2{h}(\vec{r},t)+\frac{\lambda}{2}[\nabla{h}(\vec{r},t)]^2+K(\vec{r},t).
 \eea
Here $\nu$ and $\lambda$ represent the surface tension and the
excess velocity respectively, while $K(\vec{r},t)$ is a Gaussian
noise with zero mean and co-variance
 \\$\big\langle
K(\vec{r},t)K(\vec{r'},t')\big\rangle=D\delta^d(\vec{r}-\vec{r'})\delta(t-t')$,\\
where $d$ is the dimension of the substrate, and $D$ is the noise intensity
\cite{Barabasi,6}. The interface width reads as: \bea
W(L^d,t)=\left\langle\left[\frac{1}{L^d}\sum_{\vec{r}}[h(\vec{r},t)-\overline{h}(t)]^2\right]^{1/2}\right\rangle.
\eea  This characterizes the roughness of the interface, for growth in a
substrate of length $L$, and $\overline{h}(t)$ is the spatial average
of height at time $t$. For short times, the interface scales as follows:
  \bea
  W(L,t)\approx{t^\beta}
  \label{beta}\eea
where $\beta$  is called the growth exponent. For long times, a
steady state is attained and the width is saturated as follows:
 \bea
W_{sat}(L,t)\approx{L^\alpha}.
 \label{alpha}\eea
Here $\alpha$ is the roughness exponent. Equations (\ref{beta}) and
(\ref{alpha}) correspond to limits of the dynamical relation of the Family
and Vicsek ansatz: \bea
  W(L,t)\approx{L^\alpha}{f\left (\frac{t}{L^{z}}\right)}.
  \eea
The dynamical exponent, $z=\frac{\alpha}{\beta}$, characterizes the
crossover from the growth regime to the steady state. The exact
scaling exponents are known in $d=1$, but no exact value has been
obtained in two or more dimensions \cite{11}. Many discrete models
fall into the KPZ class, such as the RSOS model \cite{kim89,kim911} and
Ballistic Deposition BD \cite{6}. %Numerical estimation of the
%scaling exponents in $d=1$ are consistent with exact values
%\cite{9,12,13} and simulations in $d=2$ are frequently used to
%estimate the corresponding exponents.
Most of the reported values of $\alpha$ range from $\alpha=0.37$ to
$\alpha=0.40$ \cite{kim89,kim911,14,15,17}, confirmed by numerical solutions
of the KPZ equation \cite{18,19,21}.

The competition between different growth mechanisms during particle
deposition, as well as phase transitions which are very often observed
in many real growth processes, has been investigated in many studies
\cite{a1,a2}.  Recently, it has been confirmed that there exists a
crossover between the Random Deposition and KPZ classes at the initial
growth stages for all  values of the height restriction parameter
between nearest neighbours for the SOS model in $1+1$-dimension
\cite{e1}. Here we are interested in investigating the
possibility of the existence of crossover in the SOS model in $2+1$
dimensions. In addition, we give a new approach to determine the
roughness exponent using Markovian properties of surfaces.

The rest of this paper is organized as follows: in section $2$, we
introduce the Markovian surface, and by using the characteristic
function, the roughness exponent is calculated. The SOS model for
finite values of $S$ is numerically investigated in section $3$.
Crossover in the growth mechanism and corresponding properties are
also investigated in detail in section $3$. Section $4$ is devoted
to conclusions and summary of our studies.
\vspace*{4mm} %\centerline{\includegraphics{fig1.eps}
\centerline{\epsfig{file=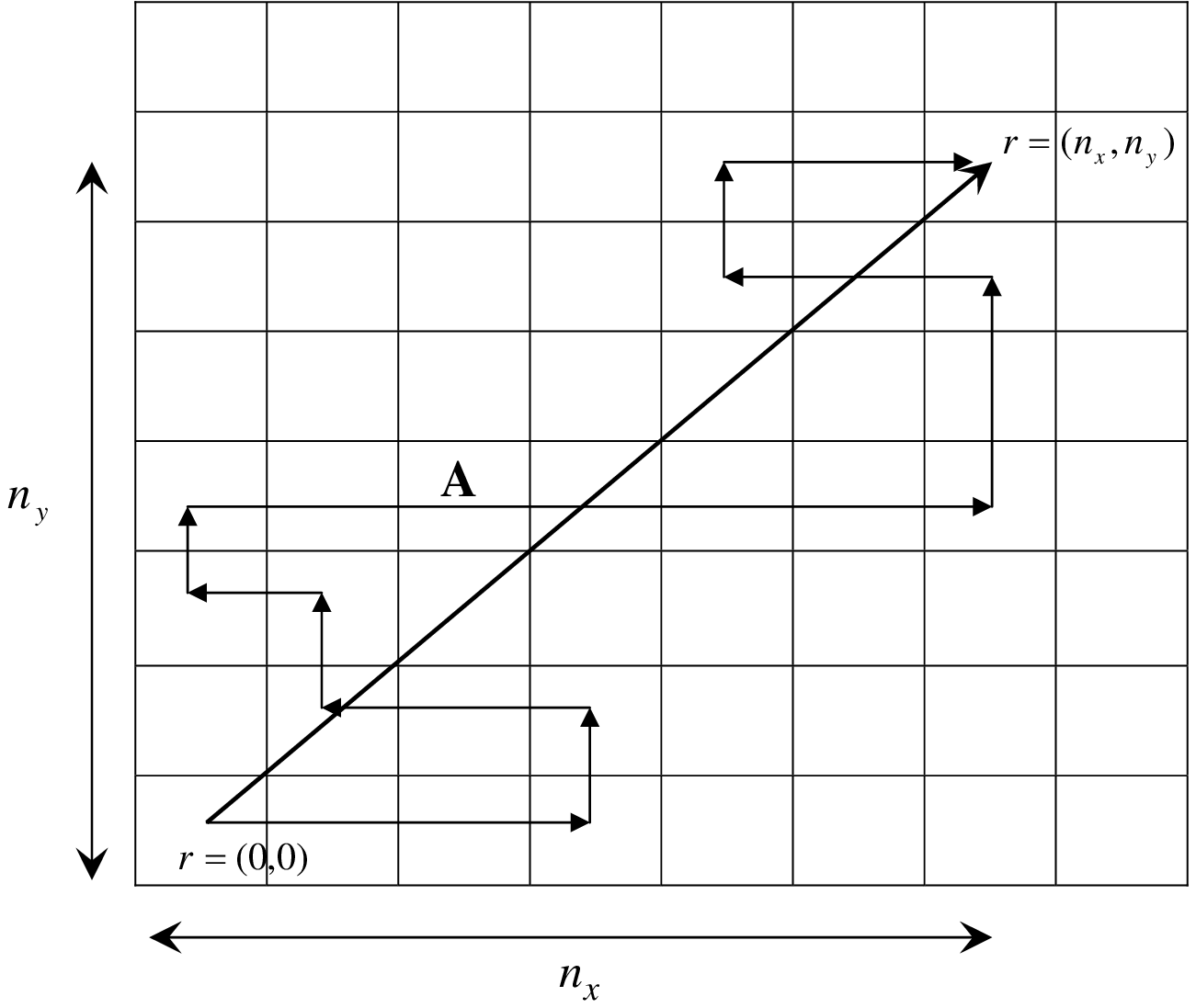,width=8cm,height=7cm,clip=}}
\begin{center}
\parbox{15.5cm}{\small{\bf Fig.1} \label{fig1}A typical trajectory from point $r=(0,0)$ to
$r= (n_x,n_y)$.}
\end{center}

%\begin{figure}[t]
%\epsfxsize=8.5truecm\epsfbox{fig1.eps} \epsfxsize=8truecm
%\narrowtext \caption{ A typical trajectory from point $r=(0,0)$ to
%$r= (n_x,n_y)$.} \label{fig1}
%\end{figure}

\section{Markovian Surface}
The Markovian surface is one of several models to represent
multi-level (stepped) crystalline surfaces. In this model, it is
assumed that the steps have only a mono-atomic height. Displacement
through any steps may be $upward$ or $downward$, each occuring with equal
probability. Let $\gamma$ be the probability of  meeting an atom
displaced vertically either upward or downward in going from any
lattice site to an adjacent one. That is,
the probability of encountering a step ($\Delta{h}=\pm 1$) while the corresponding probability for
a lateral walk, namely $\Delta{h}=0$ ($h$ is the height of the
surface), is equal to $1-\gamma$. Since every displacement or step
occurs independent of any other, the step surface is
mapped to the path of a Markovian chain \cite{22}. For the
Markovian chain or random walk model, there exist three
choices for the displacement at each walk: an upward walk with a
probability $\gamma/2$, a downward walk with a probability
$\gamma/2$ and a lateral walk with a probability $1-\gamma$
\cite{22}. As mentioned in the introduction, here we rely on the
Markovian surface to explore the scaling exponent of the RSOS growth
model. To this end, we introduce the characteristic function defined
as the Fourier transform of the probability distribution function,
$P (\Delta h(\vec{r}))$, with respect to $\Delta
h(\vec{r})=h(\vec{r})-h(0)$ after saturation time, as
   \bea Z_d(\lambda,\vec{r})=\big\langle e^{i \lambda c[h(\vec{r})-h(0)]}\big\rangle \eea
where $c$ is the unit of step variations, which is equal to one in
the Markovian surface and RSOS model. The height difference,
$h(\vec{r})-h(0)$ can be represented as the sum of the height
differences between successive sites from $r=0$ to $r=na$ in
one-dimension and to $r=\sqrt{({n_x}^2+{n_y}^2)}a$ in two-dimensions
($a$ is lattice unit). In one-dimension we have \cite{22}: \bea
h(na)-h(0)=\sum_{i=1}^n \left[h(ia)-h((i-1)a)\right]\eea For the RSOS
model in $2+1$ dimensions, the height difference between any sites
with coordinates ($n_x,n_y$) and its nearest neighbour sites
with coordinates ($n_x\pm1,n_y$ ) and ($n_x,n_y\pm1$) is
$\pm1$. To calculate the characteristic function, we should move
from point $\vec{r}=(0,0)$ to $\vec{r} = (n_x,n_y$) in different
paths like path $A$ as shown in Figure (1). So the vector sum of
the trajectories within path $A$ gives the vector $\vec{r}$. Due
to isotropy and homogeneity of the surface, the probability $\gamma$
is similar for each step. Consequently, the $Z_d(\lambda,\vec{r})$
can be written as follows: \bea Z_d(\lambda,\vec{r})\big
|_A&=&\big\langle e^{{i\lambda c[h(\vec{r})-h(0)]}} \big\rangle \big
|_A
\nonumber\\
&=&\big\langle e^{i\lambda c[h(n_x,n_y)-h(n_x-1,n_y)]} \big\rangle \nonumber\\
&&\times\big\langle e^{i\lambda c[h(n_x-1,n_y-1)-h(n_x-2,n_y)]} \big\rangle \nonumber\\
&&\times\big\langle e^{i\lambda c[h(n_x-2,n_y)-h(n_x-2,n_y-1)]}
\big\rangle\cdots \nonumber\\\eea
The number of different paths
from point $r=(0,0)$ to $r = ( n_x,n_y)$ is $(n_x+n_y-2)$ with the
condition $n_x,n_y>1$ and the number of paths from point $r=(0,0)$
to $r=( n_x,n_y)$ is $N\equiv{\bf
C}^{n_xn_y}_{n_x+n_y-1}=(n_xn_y)!/(n_xn_y-n_x-n_y+1)!(n_x+n_y-1)!$.\\
Finally the characteristic function $Z_d(\lambda,\vec{r})$ can be
written as
 \bea Z_d(\lambda,\vec{r})=Nf^{n_x+n_y-2}.\eea Here \bea f
&\equiv& \big\langle
e^{i\lambda c[h(n_x,n_y)-h(n_x-1,n_y)]}\big\rangle \nonumber\\
&=&\int d\Delta h\quad e^{ i\lambda c \Delta h(\vec{r})} P(\Delta
h(\vec{r})) \nonumber\\&=&1-\gamma[1-\cos(\lambda c)] \eea
Therefore, the characteristic function for the RSOS model is  \bea
Z_d(\lambda,\vec{r})=N\left\{1-\gamma[1-\cos(\lambda
c)]\right\}^{(n_x +n_y -2)} \label{z}\eea For $r\rightarrow 0$
regime and small $\gamma$, equation (\ref{z}) becomes:
\begin{eqnarray}
Z_d(\lambda,\vec{r})&=&e^{\gamma [1-\cos(\lambda
c)](n_x+n_y-2)}\label{zd1}
\end{eqnarray}
On the other hand, the above equation can be expressed \cite{11} as
\begin{eqnarray}
Z_d(\lambda,\vec{r})&=&e^{-\frac{1}{2}{\lambda
c}^{2}r^{2\alpha}}\label{zd2}
\end{eqnarray}
%\begin{eqnarray}
%Z_d(\lambda,\vec{r})&=&e^{-\frac{1}{2}|\phi^2|H_d(r)}\nonumber\\
%&=&\exp(-\frac{1}{2}{\lambda c}^{2}r^{2\alpha})\label{zd2}
%\end{eqnarray}
By comparing equations (\ref{zd1}) and (\ref{zd2}), one finds:
\begin{eqnarray}
r^{2\alpha}&=&\left(n_x^2+n_y^2\right)^{\alpha}\nonumber\\
&\sim& (n_x+n_y-2)
\end{eqnarray}
In order to determine the roughness exponent, $\alpha$, we have to
compute the following scaling relation for small $\vec{r}$: \bea
n_x+n_y-2\sim({n_x}^2+{n_y}^2)^\alpha \eea By using numerical
calculations and averaging over small values of $\vec{r}$, the
roughness exponent can be read as $0.39\pm0.03$ in $1\sigma$
confidence level, which is in good agreement with the previous
results \cite{kim89,kim911,14,15,17}.

\vspace*{4mm}
\centerline{\epsfig{file=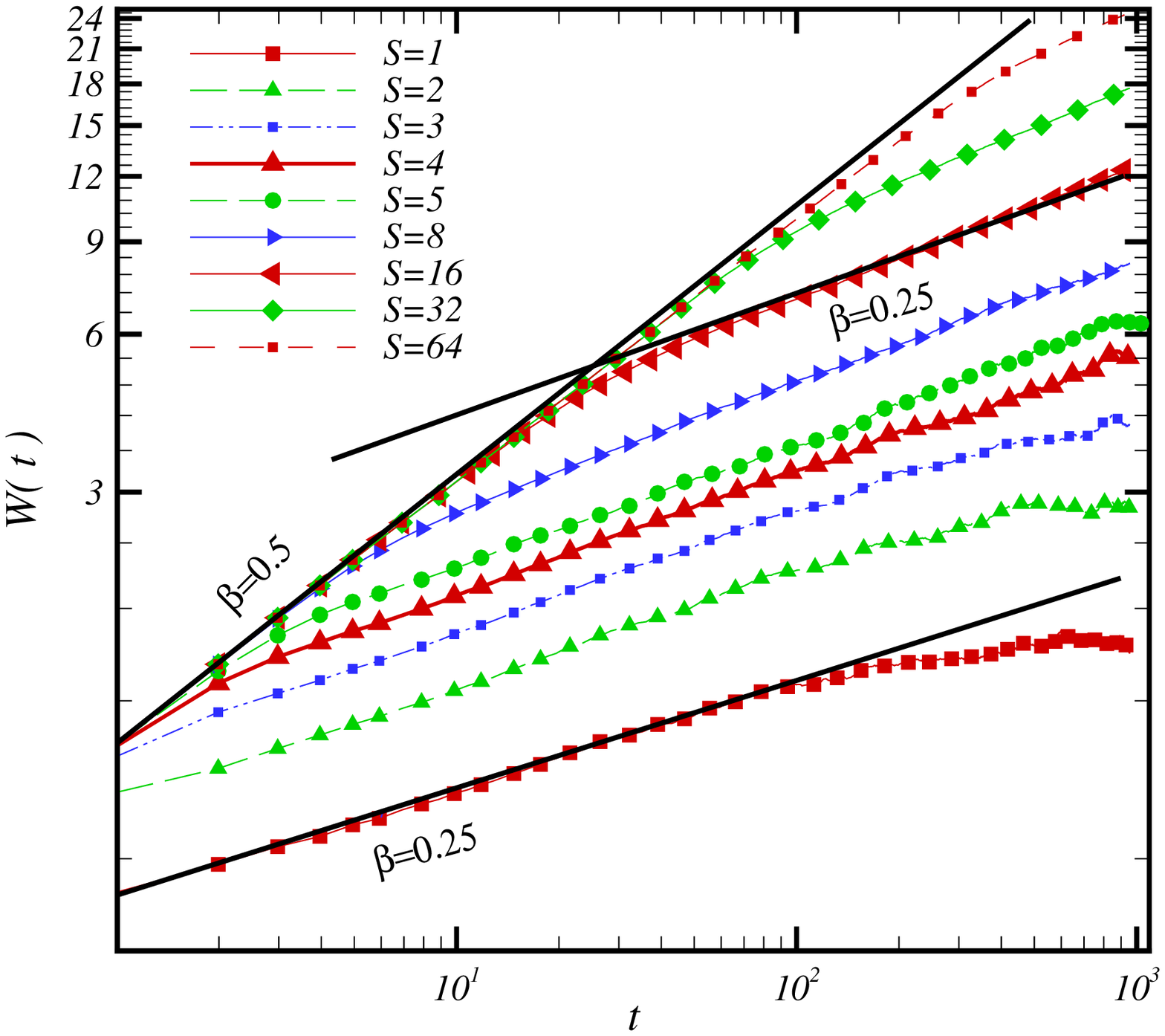,width=8cm,height=7cm,clip=}}
\begin{center}
\parbox{15.5cm}{\small{\bf Fig.2} Log-log plot of interface width versus time for
various values of parameter $S$. Here the system size is $L \times
L=4096$, and ensemble averaging has been done over $500$ independent
runs. Solid lines show the scaling behavior of RD and KPZ models in
$2+1$ dimensions with slopes $\beta=0.5$ and $\beta=0.25$,
respectively.} \label{fig2}
\end{center}

%\begin{figure}[t]
%\epsfxsize=7.9truecm\epsfbox{fig2.eps} \epsfxsize=8truecm
%\narrowtext \caption{Log-log plot of interface width versus time for
%various values of parameter $S$. Here the system size is $L \times
%L=4096$ and ensemble averaging has been done over $500$ independent
%runs. Solid lines show the scaling behavior of RD and KPZ models in
%$2+1$ dimensions with slopes $\beta=0.5$ and $\beta=0.25$,
%respectively. } \label{fig2}
%\end{figure}

\section{Simulation of Solid on Solid growth model in $2+1$ dimensions}
In the previous section, we dealt with the Markovian surface and
calculated roughness exponent for restricted solid-on-solid model in
a new way. In this section we simulate this surface growth model in
$2+1$ dimensions for finite values of parameter $S$ on the square
lattice with length $L$. The growth process of the SOS model can be described  by the following steps:\\
{\it Step 1}: Select a site randomly e.g. site $\vec{r}:(n_x,n_y)$
among all $L^2$ sites.\\
{\it Step 2}: Then all of the following conditions should be satisfied to increase the height of mentioned site:\\
 I) $|[h(n_x,n_y;t)+1]-[h(n_x-1,n_y;t)]|\leq S$\\
 II) $|[h(n_x,n_y;t)+1]-[h(n_x+1,n_y;t)]|\leq S$\\
III) $|[h(n_x,n_y;t)+1]-[h(n_x,n_y-1;t)]|\leq S$\\
IV)  $|[h(n_x,n_y;t)+1]-[h(n_x,n_y+1;t)]|\leq S$\\
V) Otherwise, do nothing. \\
{\it Step $3$}: Repeat the above tasks.\\
To reduce the errors due to the substrate's boundaries, we use
periodic boundary conditions during particle deposition. Each time
step is defined as the number of particles needed to fill the surface average
, which is equal to $L\times L$ \cite{kim89,kim911}. The
log-log plots of the $ W (L^2,t)$ versus time scale for SOS model
with different values of $S$ are shown in Figure (2). The slope of
these diagrams, for the  initial time scale of the growth process,
gives the growth exponent, $\beta $, of the model. Figure (2)
demonstrates  that there exists crossover for $t\ll t_{sat}$ in the
log-log plot of interface width versus growth time. To investigate
the crossover behavior of interface width at small time for various
values of parameter $S$, we define a fluctuation function as:
\begin{equation}
\Delta_{\diamond}(S)=\sum_t^{\tau_{\times}(S)}{|W(S,t)-W^{\diamond}_{{\rm
The}}(S,t)|}\label{delta1}
\end{equation}
where $W^{\diamond}_{{\rm The}}(S,t)\sim t^{\beta_{\diamond}}$ and
the symbol $\diamond$ stands for RD and KPZ and corresponding
exponents are $0.50$ and $0.25$, respectively. It is worth noting
that RD and KPZ are the most relevant universality classes for SOS
growth model reported in \cite{e1,kim89,kim911}. As Figure (3)
shows, by increasing the value of the restricted parameter $S$, deviation$(\delta S)$
from KPZ(RD) class at the initial growth stage increases(decreases).
After enough time, the interface width of SOS model in $2+1$
dimensions will be saturated. The value of $S$ for the intermediate
value of fluctuation function (equation (\ref{delta1})) is
$S_{\times}=8$. Subsequently one concludes the SOS model in $2+1$
dimensions for approximately $S\geq S_{\times}$ and before
saturation epoch belongs to the two different universal classes: at
the very early growth stage, it belongs to the RD class and at
intermediate time scales, $t_{\times}<t\ll t_{sat}$, it tends to the KPZ
universality class and is affected by the restrictions on the height
differences or behaves like the RSOS model with $S=1$. On the other
hand, for $S<S_{\times}$, the SOS model only belongs to one class, namely, KPZ
class, before the saturation of its interface width. The
mathematical form of this dynamic for $S\geq S_{\times}$ can be read
as \bea W(S,t)=W(S)f\left(\frac{t}{\tau_{\times}}\right)\eea in
which \bea f\left(\frac{t}{\tau_{\times}}\right)\sim \left\{%
\begin{array}{ll}
    \left(\frac{t}{\tau_{\times}}\right)^{0.50}     & \hbox{$t<\tau_{\times}$} \\
    \left(\frac{t}{\tau_{\times}}\right)^{0.25}    & \hbox{$\tau_{\times}< t\ll L^z $} \\
\end{array}%
\right.
 \eea
where $\tau_{\times}$ is the crossover time scale indicating
transition from random deposition universality class to RSOS class
which in principle depends on $S$. The quantity  $W(S)$ just
depends on $S$. Recently Ching-Chun Chien et al.\cite{e1} have shown that
the growth exponent, $\beta$, for SOS model in $1+1$ dimensions during the very early stage of surface growth is independent of $S$. Their
results confirmed that SOS model in $1+1$ dimensions for
$t<t_{\times}$ belongs to RD class and then crosses over to the KPZ
universality class for $t_{\times}<t\ll t_{sat}$ , while
the extension of their results to $2+1$ dimensions demonstrates that
for $S<S_{\times}$, it belongs to only one universality class for $t\ll
t_{sat}$ (Figure (3)).

In order to interpret the crossover in the interface width function
(see Figure (2)), we refer to the correlation function, ${\cal{C}}
(\vec{r},\vec{r'})$ defined by \bea
{\cal{C}}(\vec{r},\vec{r'})=\big\langle
[h(\vec{r},t)-\bar{h}][h(\vec{r'},t)-\bar{h}]\big\rangle \eea For
an isotropic surface, we can define the normalized correlation function
as follows: \bea C(|\vec{r}-\vec{r'}|)=\frac{\big\langle
[h(\vec{r}+\vec{l},t)-\bar{h}][h(\vec{r},t)-\bar{h}]\big\rangle}{\big\langle
[h(\vec{r},t)-\bar{h}]^2\big\rangle} \eea The correlation functions
for various values of $S$ at the early time scale, $t\ll t_{sat}$,
are shown in Figure (4). The correlation length scale, over
which the correlation function reaches  $1/e$ of its maximum,
decreases as $S$ increases. Moreover, as $S$ increases the effect of randomness in the particle deposition process decreases. This can be explained as follows : for larger values of $S$, during the very early
stage of surface growth, the height of a typical site is not
affected by its neighbouring sites due to the restriction constraint
embedded in the rule of its deposition. Consequently, one expects
the correlation length of height to decrease as $S$ increases
at the very early stage of growth. Therefore, the SOS model with
infinite $S$ reduces to the RD model during the very early growth stage. Figure (5) indicates
the log-log plot of time in which the SOS model crosses over from RD
class to KPZ class, $\tau_{\times}$ versus $S$. This confirms the
scaling behaviour of transition time versus height difference. The
slope of this plot is $\eta=1.99\pm0.02$ at $1\sigma$ confidence
interval. This value is in agreement with results in $1+1$
dimensions given by Chih-Chun Chien et al. \cite{e1}. It may be
stated that exponent has the same values in $1$ and $2$-space
dimensions, while the growth and roughness (see below) exponents
depend on space dimension.

\vspace*{4mm} %\centerline{\includegraphics{fig3.eps}
\centerline{\epsfig{file=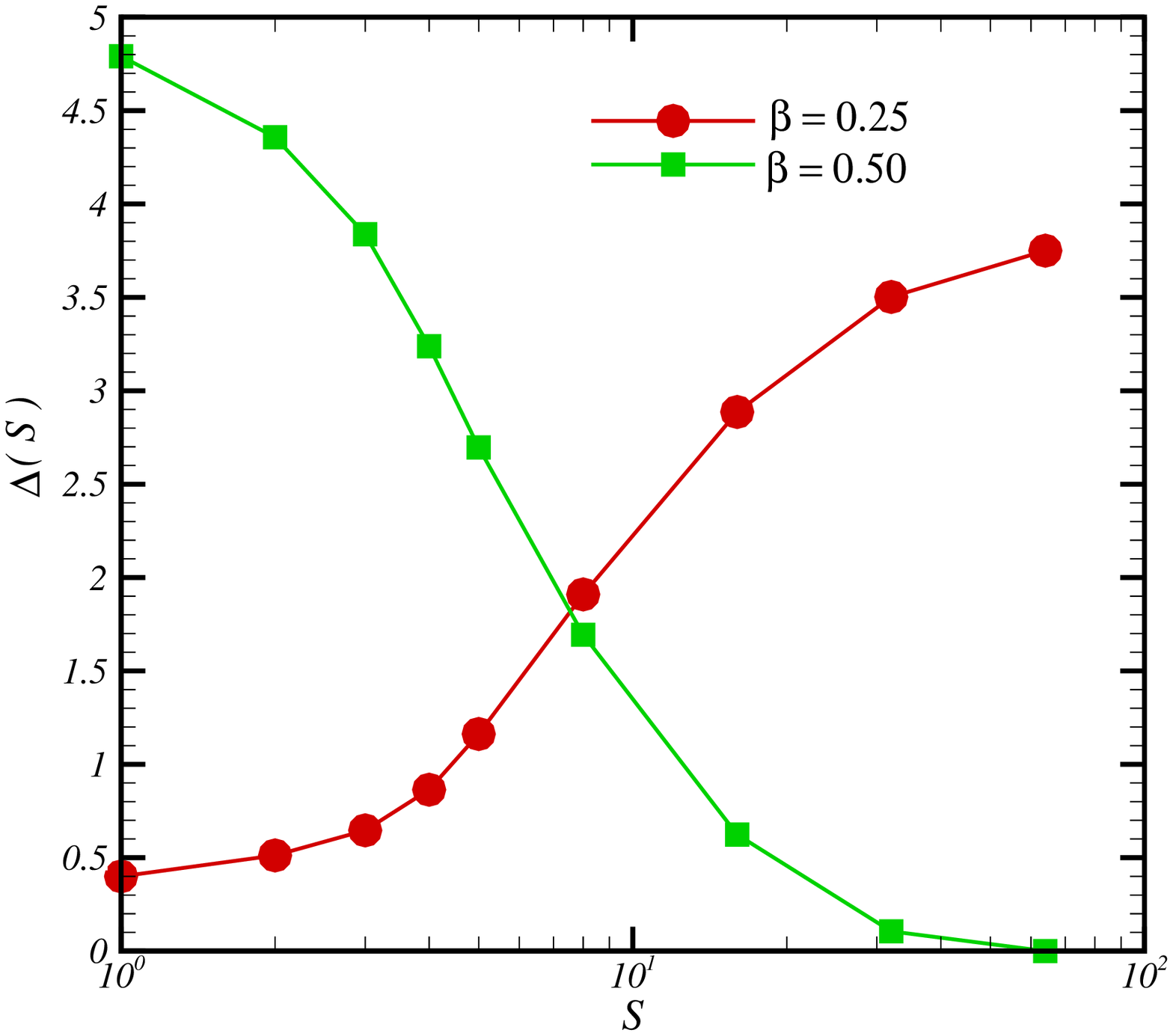,width=8cm,height=7cm,clip=}}
\begin{center}
\parbox{15.5cm}{\small{\bf Fig.3}  The function of $\Delta(S)$ for $\beta_{{\rm KPZ}}=0.25$
(filled circle symbol) and $\beta_{{\rm RD}}=0.50$ (filled square
symbol) at $t\ll t_{sat}$. } \label{delta}
\end{center}

%\begin{figure}[t]
%\epsfxsize=9truecm\epsfbox{fig3.eps} \epsfxsize=8truecm \narrowtext
%caption{The function of $\Delta(S)$ for $\beta_{{\rm KPZ}}=0.25$
%(filled circle symbol) and $\beta_{{\rm RD}}=0.50$ (filled square
%symbol) at $t\ll t_{sat}$.} \label{delta}
%\end{figure}

\vspace*{4mm} %\centerline{\includegraphics{fig1.eps}
\centerline{\epsfig{file=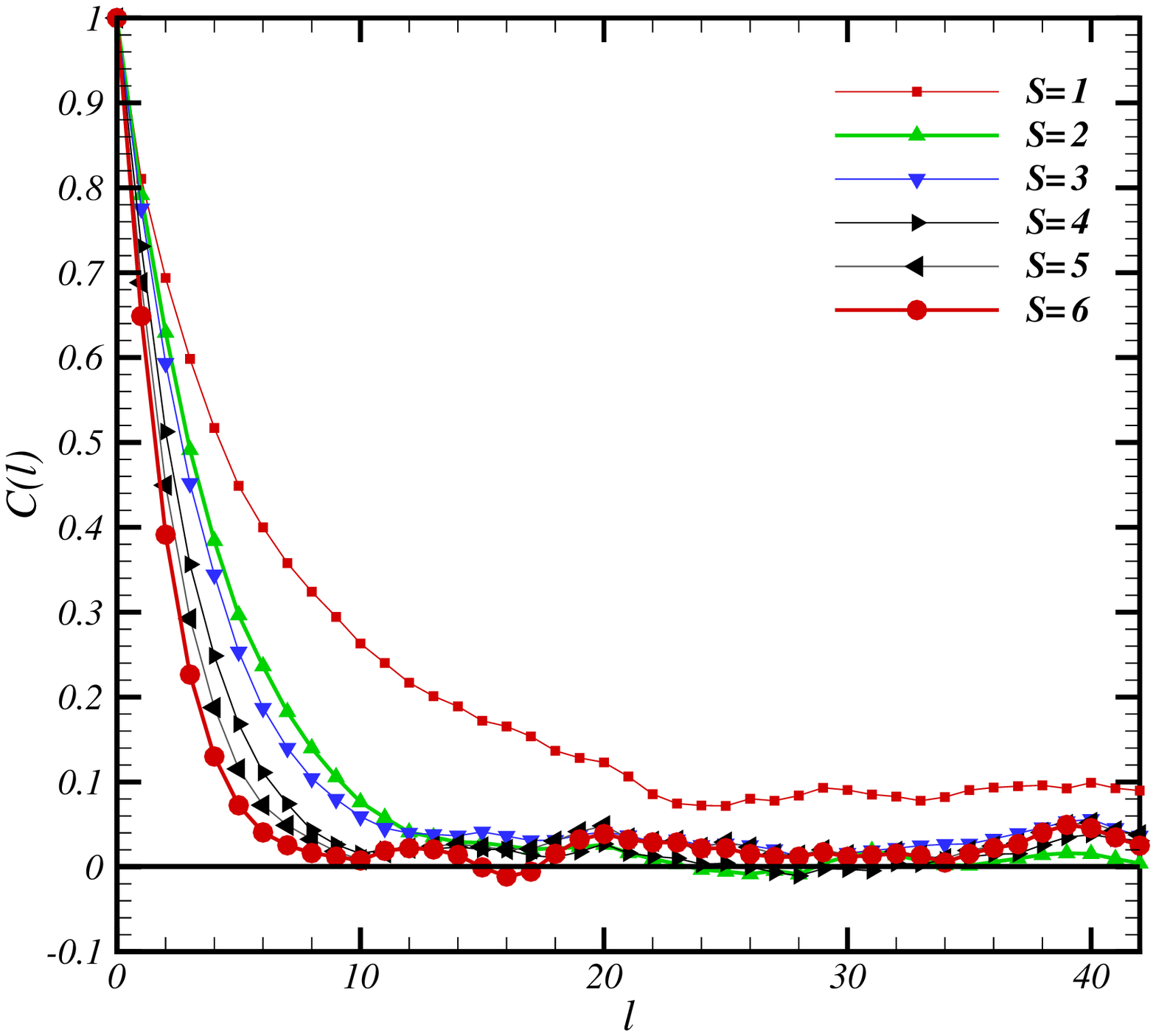,width=8cm,height=7cm,clip=}}
\begin{center}
\parbox{15.5cm}{\small{\bf Fig.4} Correlation functions for SOS model for different values of
height restriction parameters.} \label{fig6}
\end{center}
\vspace*{4mm} %\centerline{\includegraphics{fig1.eps}
\centerline{\epsfig{file=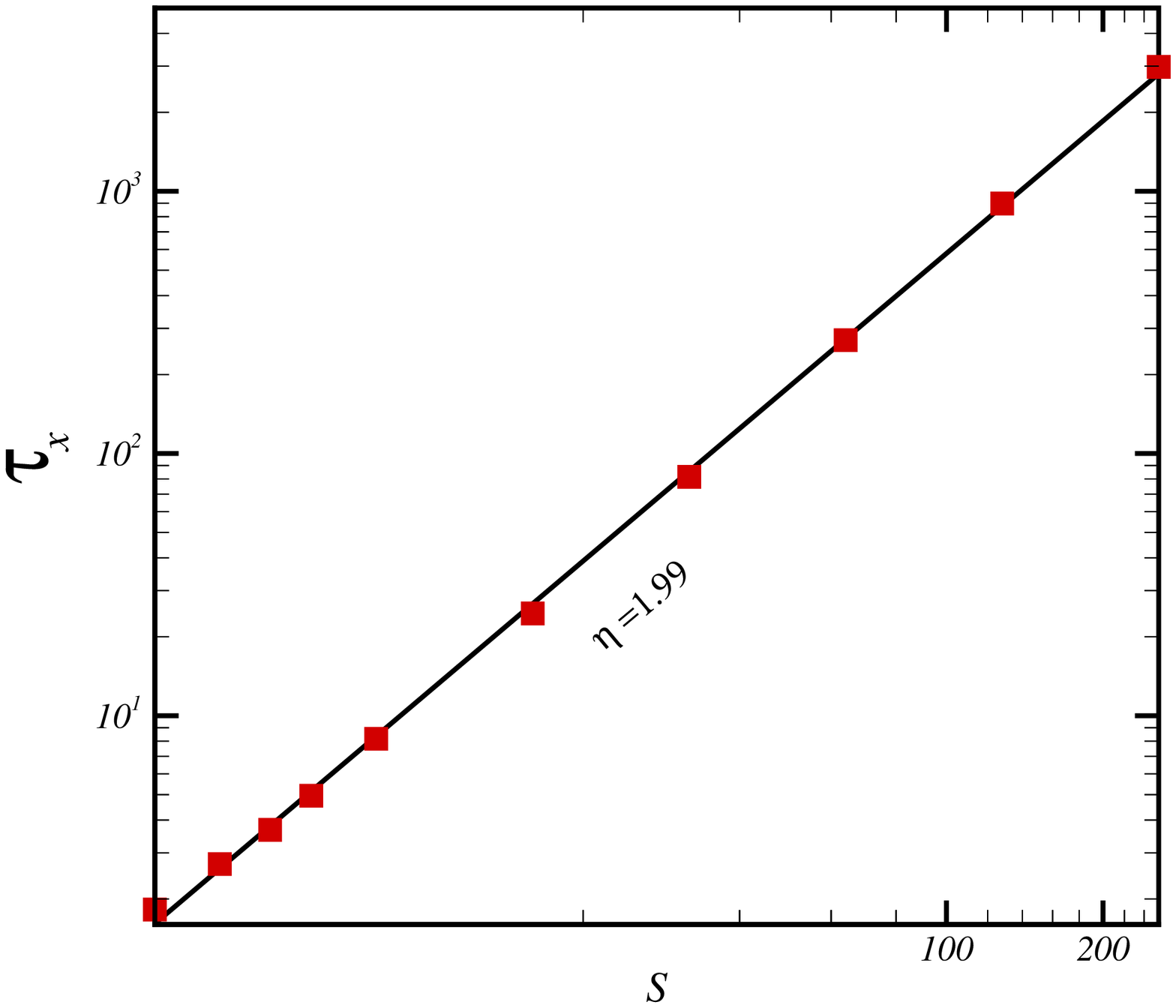,width=8cm,height=7cm,clip=}}
\begin{center}
\parbox{15.5cm}{\small{\bf Fig.5} The crossover time scale as a function of parameter $S$.
Solid line indicates the scaling function with exponent,
$\eta=1.99$.} \label{fig3}
\end{center}

%\begin{figure}[t]
%\epsfxsize=9truecm\epsfbox{fig4.eps} \epsfxsize=8truecm \narrowtext
%\caption{Correlation functions for SOS model for different values of
%height restriction parameters.} \label{fig6}
%\end{figure}
%\begin{figure}[t]
%\epsfxsize=8truecm\epsfbox{fig5.eps} \epsfxsize=8truecm \narrowtext
%\caption{The crossover time scale as a function of parameter $S$.
%Solid line indicates the scaling function with exponent,
%$\eta=1.99$.} \label{fig3}
%\end{figure}

%\begin{figure}[t]
%\epsfxsize=8truecm\epsfbox{fig6.eps} \epsfxsize=8truecm \narrowtext
%\caption{Log-log plot of height correlation function versus
%separation distance for various values of $S$. Solid line
%corresponds to scaling function with exponent, $\alpha=0.40$.}
%\label{fig4}
%\end{figure}
\vspace*{4mm} %\centerline{\includegraphics{fig1.eps}
\centerline{\epsfig{file=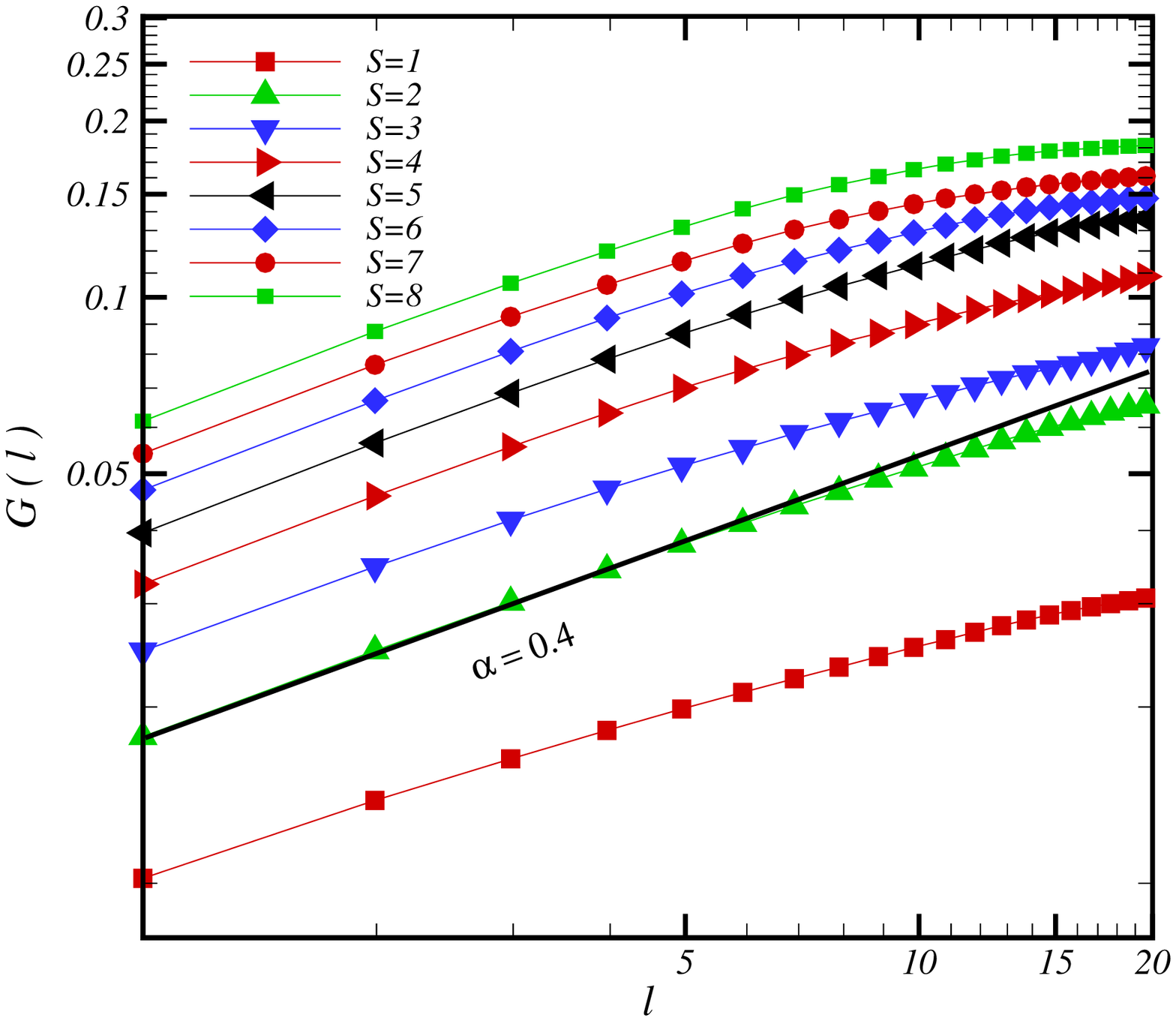,width=8cm,height=7cm,clip=}}
\begin{center}
\parbox{15.5cm}{\small{\bf Fig.6} Log-log plot of height correlation function versus
separation distance for various values of $S$. Solid line
corresponds to scaling function with exponent, $\alpha=0.40$.}
\label{fig4}
\end{center}

Another interesting parameter is the roughness exponent, $\alpha$,
which is also given via the structure function for each $S$ as
follows:
let the structure function be defined as
\bea
      G(\vec{r},\vec{r'};t)={\big\langle[h(\vec{r},t)-h(\vec{r'},t)]^2\big\rangle}^{1/2}. \eea
In the homogeneous and isotropic case $G$ depends on time scale, $t$
and separation distance, $l=|\vec{r}-\vec{r'}|$. In addition, in
principle this function may depend on $S$, so in the most general
case, the structure function reads as: \bea
G(l,S;t)=G(l,S)g\left(\frac{t}{l^{z}}\right)\eea where
$z=\alpha/\beta$. The roughness saturates after a sufficiently long
time; consequently $G (l,t>t_{sat})$ behaves as:
 \bea
G (l,S_{\rm fixed},t>t_{sat})\equiv G(l,S_{\rm fixed})\approx
l^\alpha \eea Therefore, the slope of log-log plot of $G (l,S_{\rm
fixed})$ versus $l$ for small separation distance in the saturation
regime gives the roughness exponent, $\alpha$. Figure (6) shows the
structure function for different values of $S$ in the saturation
regime. The slopes of $G(l,S_{\rm fixed})$ for small $l$ are all equal to
$0.40\pm0.01$ at $1\sigma$ confidence level which is the same as that determined for the KPZ model in
$2+1$ dimensions. This confirms that the SOS model belongs to the KPZ
universality class during the late growth stage. Figure (7) shows the log-log plot of $G(l,S)$
for small separation distance and fixed $l$ versus $S$. It
demonstrates a scaling behavior for structure function in the
saturation regime for small and fixed $l$ as a function of $S$. Its
scaling exponent is equal to $\xi=0.86\pm 0.05$, at $68\%$
confidence level. We introduce a new scaling function which gives
the relation between $l$ and $S$ after saturation epoch:
 \bea
      G(l,S)=l^{\alpha}u\left(\frac{S}{l^{z'}}\right)
      \eea
where $z'=\frac{\alpha}{\xi}$ which is a new dynamical exponent. We
also examined the height-height correlation function in $1+1$
dimensions for various values of $S$. Our results confirm a scaling
behavior with the exponent equal to $\xi=0.92\pm0.05$. Table
(\ref{Tab1}) reports all the most relevant exponents determined in
this paper as well as those given in ref. \cite{e1}.
%\begin{figure}[t]
%\includegraphics{fig7.eps} \epsfxsize=8truecm \narrowtext
%\caption{ Log-log plot of $G(l={\rm fixed})$ versus $S$ in $2+1$
%dimensions. Solid line show the scaling function with slope
%$\xi=0.86$.} \label{fig5}
%\end{figure}
\vspace*{4mm} %\centerline{\includegraphics{fig1.eps}
\centerline{\epsfig{file=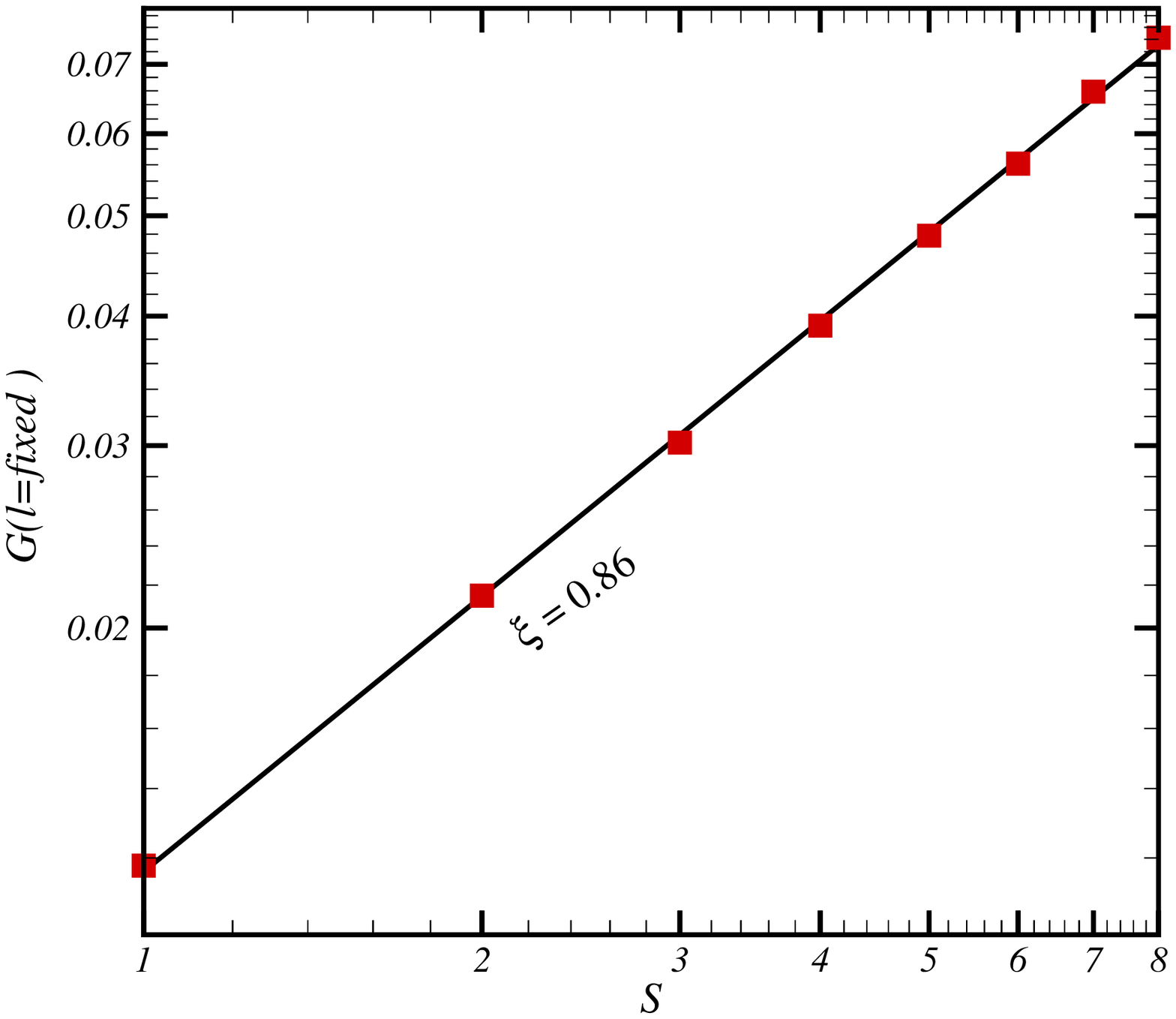,width=8cm,height=7cm,clip=}}
\begin{center}
%\vspace*{4mm}
\parbox{15.5cm}{\small{\bf Fig.7} Log-log plot of $G(l={\rm fixed})$ versus $S$ in $2+1$
dimensions. Solid line shows the scaling function with slope
$\xi=0.86$.} \label{fig5}
\end{center}
\begin{table}
\caption{\label{Tab1}Values of the scaling exponents of SOS growth
model in $1+1$ and $2+1$ dimensions.}
\begin{tabular*}{\textwidth}{@{}l*{15}{@{\extracolsep{0pt plus
12pt}}l}} \br
Dimension&$\beta$&$\eta$&$\alpha$&$\xi$&\\
\mr
&{\rm All $S$\quad[8]}&\\
$1+1$&$t<t_{\times}\quad 0.5$&$2.06$\quad[8]&$0.5$&$0.92\pm0.05$\\
&$t>t_{\times}\quad 0.33$&&
\\&&&\\&$S<S_{\times}\quad 0.25$&&\\
$2+1$&$S>S_{\times}\quad t<t_{\times}\quad
0.5$&$1.99\pm0.02$&$0.40\pm0.01$&$0.86\pm0.05$\\&$S>S_{\times} \quad t>t_{\times}\quad 0.25$&&\\
\br
\end{tabular*}
\end{table}
%\begin{table}[htp]
%\caption{\label{Tab1}Values of the scaling exponents of SOS growth
%model in $1+1$ and $2+1$ dimensions.}
%\begin{center}
%\begin{tabular}{|c|c|c|c|c|}\hline
%  Dimension&$\beta$&$\eta$& $\alpha$& $\xi$ \\ \hline&{\rm
%  All}\quad$S$\quad[8]&&&\\
%       $1+1$ & $t<t_{\times}\quad0.5$& $2.06$\quad[8]&$0.5$& $0.92\pm 0.05$   \\
%        & $t>t_{\times}\quad 0.33$&&&
%        \\\hline
%                & &&&\\&$S<S_{\times}\quad 0.25$&
%                $1.99\pm0.02$&$0.40\pm0.01$&$0.86\pm0.05$\\
%          $2+1$ &$S>S_{\times}\quads\begin{array}{cc}
%                                t<t_{\times} &0.5 \\
%                                t>t_{\times}& 0.25
%                              \end{array}$
%          &&&    \\\hline
%    \end{tabular}
%\end{center}
%\end{table}

\section{Summary and conclusion}
In this paper, we explored some scaling properties of the Solid On
Solid model, one of the well-established
algorithms for surface growth in $2+1$ dimensions. The scaling properties of
the model give deep insight into its universality
and statistical classification. To this end, we have
proposed a new method for finding the roughness exponent for one of
the special classes of the SOS model: the so-called RSOS model. In the RSOS model, the value of near-neighbours height difference is restricted to $S=1$. This new method uses
Markovian surfaces properties. By introducing a characteristic
function we computed the scaling behavior of the site's numbers with
respect to the various values of the displacement vector. The value of the
roughness exponent derived by the Markovian approach is
$\alpha=0.39\pm0.03$ at $68.3\%$ confidence interval. Our result for the
roughness exponent is in good agreement with those obtained from
direct simulation of the RSOS model performed in earlier works
\cite{kim89,kim911,14,15,17}. We simulated the growth surface according
to the SOS algorithm explained in section $3$ for finite values of $S$.
The interface width as a function of time has a crossover time scale
for approximately $S\geq 8$ which confirms two universality classes
for the SOS model at the very early stage of its growth. The SOS model
for $S\geq 8$ at $t<\tau_{\times}$ falls under the random deposition
universality class and thereafter tends to the KPZ universality class for
$t_{\times}<t\ll t_{sat}$. To make these crosses over more obvious, we
investigated the normalized correlation function. Our results
indicate that by increasing the height difference parameter at
$t\ll t_{sat}$, the correlation length scale is decreased (see
Figure (4)). This is clearly due to the restriction constraints
which eliminate the effect of neighbors on the memory of particle
deposition. With increasing growth time, particle depositions are affected by the
restriction rule; consequently, the SOS model with infinite $S$ tends to
the KPZ growth model before the saturation time stage.
 According to
Figure (5), the scaling exponent of $\tau_{\times}$ versus $S$ is
 $\eta=1.99\pm0.02$ which is in agreement with that
obtained in $1+1$ dimensions  \cite{e1}, while the growth and
roughness exponents depend on the space dimension. In order to compute
the roughness exponent, we used a structure function. The slope of
this structure function versus distance separation, $l$ in the log-log
scale for various values of parameter $S$ is $\alpha=0.40\pm0.01$ at
$68\%$ confidence level. This quantity, in contrast to the growth exponent,
did not show crossover behaviour in the log-log plot of structure
function as a function of separation distance (see Figure (6)). The
value of the roughness exponent also confirms our result regarding the
universality class of the SOS model at longer times. Based on Figure (7),
we found that the height-height correlation function, $G(l)$ for
small and fixed $l$ versus $S$ indicates a scaling behavior with
exponent $\xi$ which is equal to $0.86\pm0.05$ and $\xi=0.92\pm0.05$
at $1\sigma$ level of confidence for $2+1$ and $1+1$ dimensions,
respectively. Using the roughness and $\xi$ exponents we introduced
a new dynamical exponent as $z'=\alpha/\xi$.

{\bf Acknowledgements} We thank Ronald Begg and Tony Anderson for carefully reading the manuscript and providing us with useful comments. M. Sadegh Movahed is grateful to the School
of Astronomy and Astrophysics, Institute for Research in Fundamental
Sciences (IPM), Tehran, IRAN for its hospitality during the
preparation of this manuscript.


\begin{thebibliography}{10}

\bibitem{e3}A. Pimpinelli, {\it Physics of crystal growth }, Cambridge University Press, Cambridge, 1998
\bibitem{a3}A. Kolakowska, M. A. Novotny and P. S. Verma, Phys,
Rev. E 73 (2006)011603.
\bibitem{a4}C. M. Horowitz and E. V. Albano, Eur. Phys. J. B 31,
(2003)563-569.
%\bibitem{a5}Egypt. J. Sol. v 25, (2002) N 2
\bibitem{a6}Yup Kim, S. Y. Yoon and Hyunggyu Park, Phys. Rev E 66
(2002) 040602
\bibitem{2} F. D. A. Aarao Reis, Phys. Rev. E 69, (2004)021610.
\bibitem{3} {\it Frontiers in Surface and Interface Science}, edited by Charles B.Duke and E.Ward Plummer(Elsevier, Amesterdam, 2002)
\bibitem{Barabasi} A . L. Barabasi, H. E. Stanly, {\it Fractal Consepts in Surface Growth}, Cambridge Universty Press, New York, 1995.
\bibitem{e1}Chih-Chun Chein, Ning-Ning-Pang, Phys. Rev. E {\bf 70},(2004) 021602 .
\bibitem{e2}T. Halpin-Healy  and Y.-C.Zhang, Phys. Rep. 244,(1995) 215
\bibitem{4} F. D. A. Aarao Reis, Physica A 316(2002)250-258.
\bibitem{5}F. Family, J. Phys. A 19(1996)L441.
\bibitem{6}M. J. Vold, J.Colloid, Sci. 14(1959)168.
\bibitem{7} F. Family, T. Vicsek, J Phys. A 18(1958)L75.
\bibitem{8}R. Jullien and R. Botet, J. Phys. A 18,(1985)2279.


\bibitem{famili90} F. Family, Physica A 168, (1990)561 .
\bibitem{11}J. Krug, Advances Phys. 46,(1997) 139 .
\bibitem{kim89}Jin Min Kim and J. M. Kosterlitz, Phys. Rev. Lett. 62,
19(1989)2289-2292  % 9
\bibitem{kim911} J M Kim, J M Kosterlitz and T Ala-Nissila, J. Phys. A: Math. Gen. 24, 5569-5586(1991).
\bibitem{sarma} S. Das Sarma, J. Vac. Sci. Technol, A {\bf 8}, 2714
(1990); B {\bf 10}, (1992)1695 .
\bibitem{muller05}H. M$\ddot{{\rm u}}$ller-Krumbhaar, F. Gutheim and C. P$\ddot{{\rm u}}$tter, Journal of Crystal Growth 275 (2005)
51–55.

\bibitem{margo90} A. Margolina 1'2 and H. E. Warriner, Journal of Statistical Physics, Vol. 60,(1990) Nos. 5/6,
\bibitem{amar90}J. G. Amar and F. Familiy, Phys. Rev. Lette. {\bf{64}}, no. 5, p.   543 (1990).
\bibitem{e4}Z. Ding, D. W. Bullock, P. M. Thibado and K. Mullen, Phys. Rev.
Lett. 90, (2003)216109.
\bibitem{zhd03} V. P. Zhdanov and B. Kasemo, Surf. Sci. 418,(1998) 84 .
\bibitem{a1}T. J. Oliveira, K. Dechhom, J. A. Redinz and F. D. A. Aarao
Reis, Phys. Rev. E 74, 011604 (2006).

\bibitem{a2}F. D. A. Aarao Reis, Phys. Rev. E 73, (2006) 021605.
\bibitem{a7}F. D. A. Aarao Reis, Physica A 316 (2002) 250-258
\bibitem{24}S. Kimiagar, G. R. Jafary, M. R. Rahimitabar, J. Stat. Mech. (2008) P02010
\bibitem{sarma92}, S. Das Sarma and S. V. Ghaisas, Phy. Rev. Lett.,
69, 26 (1992).
\bibitem{ala92} T. Ala-Nissila, T. Hjelt and J. M. Kosterlitz,
Europhys. Lett., 19 (1),(1992) 1-5 .
\bibitem{lee}Sang Bub Lee, Hyeong-Chai Jeong and Jin Min Kim, J.
Stat . Mech. (2008) P12013

\bibitem{park95} K. Park and B. N. Kahng, Phys. Rev. E 51,(1995) 796 .
\bibitem{huang98}Z. F. Huang and B. L. Gu, Phys. Rev. E 57,(1998) 4480 .
\bibitem{10}M. Kardar, G. Parisi and Y. C. Zhang,  Phys. Rev. Lett. 56,(1986)889.
\bibitem{14}P. Devillard and H. E. Stanley, Physica A 160, (1989)298.
\bibitem{15}D. Liu and M. Plischke, Phys. Rev. B 38, (1988)4781.
\bibitem{17}T. Ala-Nissilia, J. Stat. Phys. 72, (1993)207.
\bibitem{18}J. G. Amar and Family, Phys. Rev. A 41, (1990)3399.
\bibitem{19}K.Moser, Physica A 178,(1991) 215.
\bibitem{21}M. Lassig, Phys. Rev. Lett. 80,(1998)2366.
\bibitem{22}H. N. Yang, G. C. Wang, {\it Diffraction From Rough Surfaces and Dynamic Growth Fronts},World Scientific
Publishing, 1993.







\end{thebibliography}
\end{document}